\documentclass[%
 preprint,
 amsmath,amssymb,
 aps,
]{revtex4-1}
	
\usepackage{graphicx}% Include figure files
\usepackage{dcolumn}% Align table columns on decimal point
\usepackage{bm}% bold math
\begin{document}
	
	\preprint{APS/123-QED}
	
	\title{Spider silks mechanics: \\ predicting humidity and temperature effects}% Force line breaks with \\
	\author{Giuseppe Puglisi}%
	\email{giuseppe.puglisi@poliba.it}
	\affiliation{%
		Department of Civil Environmental Land Building Engineering and Chemistry, Polytechnic University Bari, via Orabona 4, 70125 Bari, Italy.	}%

	\author{Domenico De Tommasi}
	\affiliation{Department of Civil Environmental Land Building Engineering and Chemistry, Polytechnic University Bari, via Orabona 4, 70125 Bari, Italy.}%Lines break automatically or can be forced with \\
	
	\author{Vincenzo Fazio}
	\affiliation{Department of Civil, Environmental and Mechanical Engineering, University of Trento, Via Mesiano 77, 38123 Trento, Italy.}%
	
	\author{Nicola Maria Pugno}
	\email{nicola.pugno@unitn.it}%
	\affiliation{%
		Laboratory of Bio-inspired, Bionic, Nano, Meta Materials and Mechanics, Department of Civil, Environmental and Mechanical Engineering, University of Trento, Via Mesiano 77, 38123 Trento, Italy;\\ School of Engineering and Materials Science, Queen Mary University of London, Mile End Road, London E1 4NS, UK.
	}%

	\date{\today}% It is always \today, today,
	
\begin{abstract}
	We deduce a microstructure inspired model for humidity and temperature effects on the mechanical response of spider silks, modelled as a composite material with a hard crystalline and a soft amorphous region. Water molecules decrease the percentage of  crosslinks in the softer region inducing a variation of natural configuration of the macromolecules. The resulting kinematic incompatibility between the regions crucially influences the final mechanical response. We demonstrate the predictivity of the model by quantitatively reproducing the experimentally observed behavior.
	\end{abstract}
	
	%Use showkeys class option if keyword
	\maketitle
	
	%\tableofcontents
	
Due to their extraordinary properties, spider silks represent one of the most intensively studied materials, also in the spirit of biomimetics  \cite{ZNW}. The availability of detailed experimental analyses let in the last decades a deeper understanding --both from a chemical and structural point of view-- of the complex multiscale system at the base of their notable mechanical behavior. Nevertheless, many important phenomena regulating its loading history and rate dependence together with temperature and humidity effects, remain unclear \cite{review}. 
The striking phenomenon we focus on in this paper is the so called \textit{supercontraction} effect. Firstly addressed in \cite{Work_1977}, it consists in a shrinkage of the fiber up to 50\% of its initial length, when immersed in water or in high humidity environment strongly modifying the mechanical performances. 
	
At the molecular scale, spider silk is composed by an amorphous matrix of oligopeptide chains and by pseudo-crystalline regions made up principally of polyanaline $\beta$-sheets \cite{elices_hidden_2011, sponner_composition_2007} with dimensions between $1$ and $10$ nm \cite{keten_nanostructure_2010}, mostly oriented in the fiber direction \cite{jenkins_characterizing_2013}. 
The cross section of the fiber is highly organized in the radial direction \cite{li-new-1994,eisoldt-decoding-2011,sponner_composition_2007}. Moreover, the chemical and structural composition varies according with the different silks produced by the different glands \cite{Cranford_chapter12} and of course the different species. Here, to fix the ideas, we focus on the most performant case of dragline silk. 
		
\begin{figure}[t!] \vspace{-0.5 cm}
	\hspace{-0.8 cm}  	\includegraphics[width=.55\textwidth]{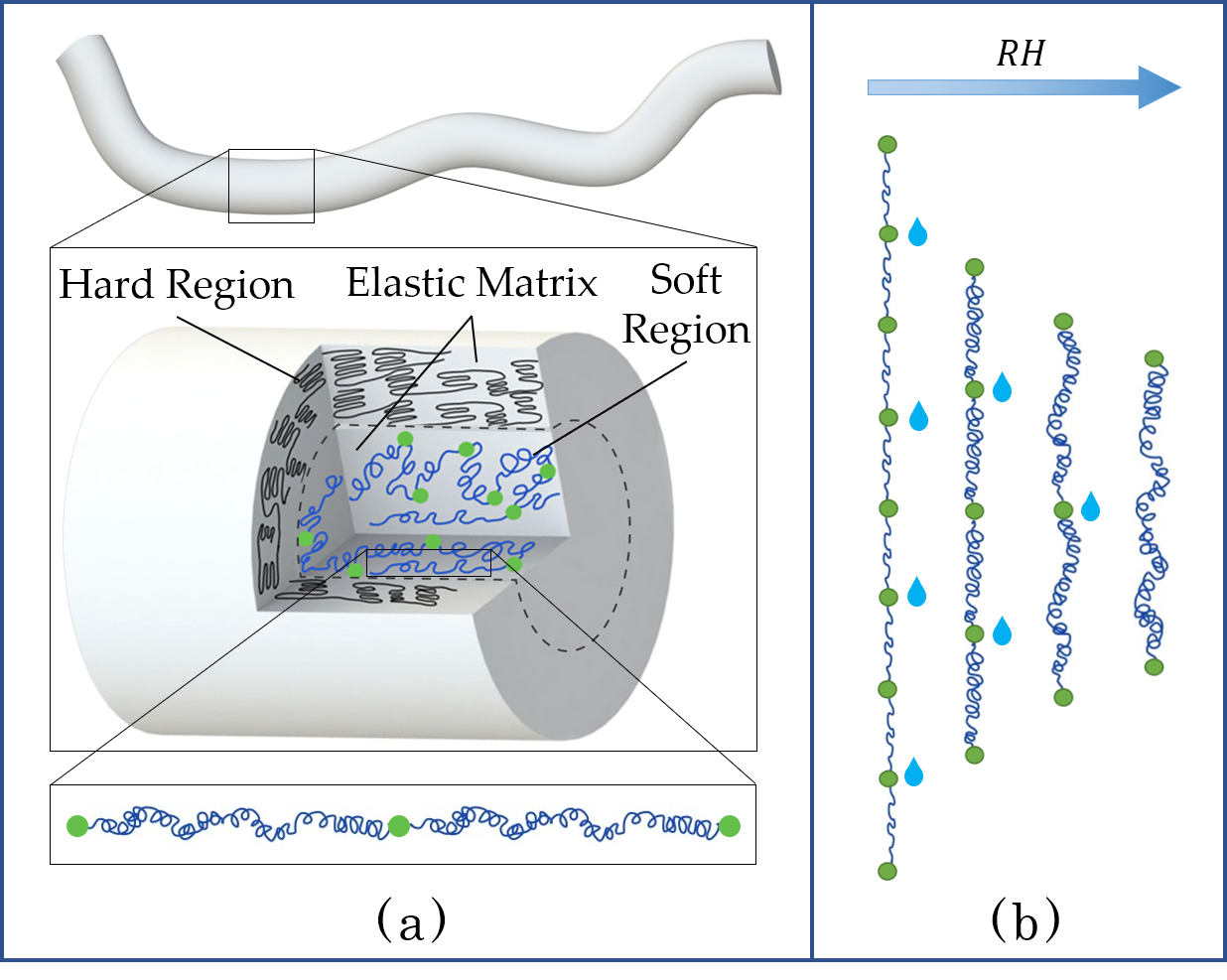}\hspace{-0.85 cm} \vspace{-0.2 cm}
	\caption{\label{fig:scheme_composition_3d} Cartoon of the silk fiber and its  microstructure (a). The outer hard region is represented by black chains, the soft region molecules are drafted in blue, whereas the  embedding elastic matrix is grey. In (b) we schematize the effects of water molecules disrupting crystal domains (green dots) and inducing entropic chain recoiling. }\vspace{-0.4cm}\end{figure}

More in detail, the thread is covered by a skin, with a chemical and physical protection function, that does not play a role in supercontraction and mechanical response~\cite{yazawa_role_2019}. We thus neglect it in the model. Next, the core can be  schematically decomposed as in Fig.~\ref{fig:scheme_composition_3d}.
The major constituent of the external part \cite{li-new-1994,brown_critical_2011}  are proteins (Major ampullate Spidroin 1, MaSp1) organized into $\beta$-pleated sheets. We refer to this fraction as {\it hard region}.
The internal part, here referred as {\it soft region}, is instead mainly constituted by proteins with a proline content preventing the formation of $\beta$-sheet structures \cite{sponner_composition_2007} (Major ampullate Spidroin  2, MaSp2). This fraction has a significantly lower crystallinity and macromolecules with  weaker crystal domains, typically in the form of $\alpha$-helix and $\beta$-turns \cite{sponner_composition_2007,nova_molecular_2010}. 
The different crystallinity is also due to the shear stress at the spinning duct wall inducing the formation of  $\beta$-sheets in the outer region \cite{giesa_2016,brown_critical_2011}.

\vspace{0.2 cm}

Based on the previous description and referred literature, we model the silk fiber as a composite material with a hard external fraction of crystalline chains and a soft internal fraction of amorphous chains. Moreover, by following the classical approach for polymeric and biopolymeric materials \cite{flory_1982}, we consider a tridimensional elastic matrix that embeds hard and soft fractions, describing the complex macromolecular network composing the spider thread, with inner and intrachains connections. 

Due to the different crystalline composition, the humidity affects differently the hard and soft fraction. Water hardly breaks the H-bonds of the compact  $\beta$-sheet domains in the hard fraction \cite{yazawa_role_2019}. On the other hand, here, we may observe a misalignment of the crystals with respect of the fiber direction that increases as the relative humidity {\small RH} grows \cite{eles_strain_2004}. Since the material stiffness grows with the alignment of the crystals~\cite{du_design_2006}, water induces a humidity dependent damage that we introduce in our model. On the contrary, water content strongly influences the crystal percentage in the soft internal core \cite{du_design_2006,elices_finding_2005,elices_hidden_2011}, because $\alpha$-helices and $\beta$-turns are much more easily broken by water molecules.  In particular, the experiments exhibits a non uniform variation of the silk properties with a localized transition at a specific value of {\small RH}, hereon indicated as {\small RH}$_c$, known as supercontraction threshold \cite{fu_moisture_2009}. Finally,
an important effect in the evolution of the natural configuration is induced from the stretch history \cite{puglisi_micromechanical_2017}. Indeed, as the end-to-end molecule length changes, $\beta$-sheets undergo unravelling with a corresponding increase of the number of available monomers, here considered in the hard fraction.

\section{ Micromechanical model} 
\noindent According with classical Statistical Mechanics results \cite{rubinstein2003polymer,cohen_origin_2021}, the expectation value of the end-to-end length for ideal chains is 
\begin{equation}\label{SM}L_n=< r^2 >^{1/2}= b\, n^{1/2}, \end{equation} 
where $n$ is the number of Kuhn segments with length $b$. As described above, $n$ depends on  humidity in the soft fraction, whereas it depends from the maximum attained stretch in the hard fraction.\vspace{0.2 cm}
\hspace{-0.55 cm}

\subsection{Soft Region.}  To consider the disruption process of H-bonds induced by hydration  \cite{du_design_2006}, let us introduce the function $m=\hat m(\mbox{\small RH})$ assigning the number of links in the generic humidity state, with $m(0)=m_o$ and $m(100)=m_f$ (initial and permanent number of H-bonds) \cite{vollrath_spider_2006}.  To the knowledge of the authors, no direct measurement of $\hat m$ is available, so that we consider a Gaussian probability density of rupture events
\begin{equation} \label{eqn:drh}
	d(\text{\small RH})=\frac{\hat m(\mbox{\small RH})-m_o}{m_f-m_o}= \int_0^{\mbox{\tiny RH}} \frac{1}{\sqrt{2\pi s ^{2}}}\;e^{-{\frac {\left(\mbox {\tiny RH}-\text{\tiny RH}_c \right)^{2}}{2s ^{2}}}}. \end{equation}
Here $d\in (0,1)$ is a `damage' type parameter, measuring the percentage of broken links and we assume that the Gaussian is centered in the critical value {\small RH}$_c$ (see details and Fig.~A1 in the Appendix). 

To obtain the corresponding variation of the natural length based on \eqref{SM},
assume that  $n_o^s$ is the (mean) number of chain free monomers when the silk is spun. Here and in the following we indicate by the apexes $s, h, m$, and $t$ the soft,  hard, matrix and homogenized (total) quantities. If we identify the number of H-bonds with the number of domains in which the chain is divided (see the scheme Fig.~\ref{fig:scheme_composition_3d}(b)), the mean number of free monomers in each domain is  $n^s=\hat n^s(\text{\small RH})=n_o^s/\hat m(\text{\small RH})$ corresponding to a natural length	
\begin{equation} \label{eqn:lncs}
		L^s_n=\hat m(\text{\small RH})\sqrt{n_o^s/\hat m(\text{\small RH})}\, b^s=\sqrt{n_o^s\, \hat m(\text{\small RH})}\, b^s. \end{equation}
We obtain in this way an analytic measure of the shrinkage chain effect induced by humidity.
Observe that instead the contour length is fixed: $L^s_c=n^s\, b^s.$ \vspace{0.2 cm}

 \subsection{Hard Region} 
 \noindent As anticipated, following~\cite{du_design_2006} we assume that  the elastic modulus of the crystalline region decreases with {\small RH}  by considering a (phenomenological) damage function (see Eqn.~\eqref{eqn:ehb}). 
On the other hand, while $\beta$-sheet crystals are affected only in their orientation by humidity, large strain can induce important unravelling effects as fully described in \cite{puglisi_micromechanical_2017} with conformational transitions inside the secondary structure \cite{Cranford_chapter12,giesa_2016}, from a coiled configuration \cite{yarger_uncovering_2018} to an unfolded state \cite{Cranford_chapter12,puglisi_micromechanical_2017} (see the scheme in Fig.~A3 in the Appendix).
Thus the mean number $n^h$ of available free monomers depends on the maximum attained value of the end-to-end length: $n^h=\hat{n}^h(L^h_{max})$:		
\begin{equation} \label{eqn:lnch}
		L^h_n=\sqrt{\hat n^h(L_{max})} \ b^h, \qquad	L^h_c=\hat n^h(L_{max})\, b^h. \end{equation} 
Observe that for simplicity we assume that the unfolding is irreversible with $n^h$ monotonically increasing with $L_{max}$. More general hypotheses could be introduced~\cite{de_tommasi_damage_2010}.	
\vspace{0.2 cm}

\section{From single chain to macro laws}

\noindent Both in the amorphous and crystalline region we adopt the Worm Like Chain (WLC)  energy in the form proposed in \cite{de_tommasi_energetic_2013} $\varphi_e=\varphi_e(L,L_c)=\kappa\frac{L^2}{L_c-L}$ where $\kappa=\frac{k_B T}{4 l_p}$, $T$ is the temperature, $k_B$ the Boltzmann constant and $l_p$ the persistent length \cite{rubinstein2003polymer}. This  energy respects the limit extensibility condition, $\lim_{L \rightarrow L_c} \varphi_e(L,L_c)=+\infty$, and allows for explicit calculations. Moreover,
following \cite{TPD},  we extend this function to consider that, as described above, the end-to-end length $L$ can be decomposed in the permanent part measured by \eqref{SM}  and an elastic part $L_e= L-L_n$. Thus we assume $\varphi_e=\kappa\frac{L_e^2}{L_c-L}$ and a force-elongation law
\begin{equation} \label{eqn:force-stretch}
	f=\frac{\partial\varphi_e}{\partial L}=
	\kappa \left[\left( \frac{L_c-L_n}{L_c-L}\right)^2-1 \right]. \end{equation}
Observe that the force decreases to zero when the length attains its natural value ($L=L_n$ or $L_e=0$).
	
We remark that the proposed model can be inscribed in the theory of Thermodynamics with internal variables~\cite{Coleman-1967} in the simple case when there is a single external variable $L$ and a single internal variable  $L_{max }$. In our simple setting of isothermal processes, to verify the thermodynamic consistency of the model we consider the Clausius-Duhem inequality, requiring the positivity of the  dissipation rate $\Gamma =f \dot L - \dot \varphi_e (L,L_{max}) \geq 0$.
Since at given $\text{\small RH}$ the only material fraction involved in the dissipation is the hard one, undergoing unfolding effects and variations of the natural length regulated by $L_{max}$ according with \eqref{eqn:lnch}, the internal energy dissipation rate reduces to $Q'(L_{max})=-\partial_	{L_{max}} \varphi_e(L,L_{max})$. 
Thus, in view of \eqref{eqn:force-stretch}, we obtain $\Gamma=Q'(L_{max})\dot L_{max}= \frac{L^2}{(L_c(L_{max})-L)^2}  \frac{d \hat n(L_{max})}{d L_{max}} b^h\geq 0$ that is satisfied under our assumption that $\hat n^h$ is  increasing.
	
Eventually, to obtain the macroscopic behavior of the thread we consider  the classical \textit{affinity hypothesis} \cite{rubinstein2003polymer} that identifies the macroscopic stretches with the macromolecular ones. 
We can then introduce the following stretch measures of the different considered fractions 
\begin{equation} \label{eqn:stretch measures}
	\begin{tabular}{l}
		$\lambda^i=\frac{L}{L^i_o}$ \quad \small\text{total stretch,}	\\[6pt]
		$\lambda^i_e=\frac{L^i_e}{L^i_o}$ \quad \small\text{elastic stretch,}	\\[6pt]
		$\lambda^i_n=\frac{L^i_n}{L^i_o}$ \quad \small\text{permanent stretch,}		\\[6pt]
		$\lambda^i_c=\frac{L^i_c}{L^i_o}$ \quad \small\text{contour stretch,}
	\end{tabular}
	\quad i= h,s,m,t, \end{equation}
with $L^i_o=b^i\, \sqrt{n_0^i}$ the initial natural length. 
	
%soft region
The natural and contour stretches for the soft region can be deduced using Eqns.~\eqref{eqn:drh}, \eqref{eqn:lncs} and \eqref{eqn:stretch measures} (see details and Fig.~A2 in the Appendix):
\begin{equation} \label{eqn:lanbb}
	\lambda^s_n=\sqrt{\frac{\hat m(\text{\small RH})}{m_o}}=\sqrt{1+d\left( \frac{m_f}{m_o}-1 \right) }, \hspace{0.1cm} 	\lambda^s_c=\sqrt{\frac{n_o^s}{m_o}}	\end{equation}
Under an additive assumption and given  the number of chains per unitary reference area $N^s_{fib}$, the (Piola, engineering) stress using Eqns.~\eqref{eqn:force-stretch} and \eqref{eqn:stretch measures} is given by
\begin{equation} \label{eqn:sigma stretch soft part}
	\sigma^s=E^s \left[ 
	\left( \frac{\lambda^s_c-\hat{\lambda}^s_n(\text{\small RH})}{\lambda^s_c-\lambda^s}\right)^2-1
	\right], \end{equation}
where the natural and contour stretches are given by Eqn.~\eqref{eqn:lanbb}, whereas $E^s=N^s_{fib} \kappa^s$ is the elastic modulus of the soft fraction. 
	
%hard region
For the hard region (see the Appendix) the contour and natural stretches are calculated by using Eqns.~\eqref{eqn:lnch} and \eqref{eqn:stretch measures}, with $L^h_o=\sqrt{n^h_o}\, b^h$ the initial natural length and $n^h_o$ the initial mean number of available free monomers. We have 
$\lambda^h_c=\hat{n}^h (\lambda_{max})/\sqrt{n^h_o}$ and $\lambda^h_n
	=\sqrt{\hat{n}^h (\lambda^h_{max})/n^h_o}$, respectively. 
Since the effective stretch-induced unfolding depends on the unknown size and strength crystals distribution, following \cite{TPD} we assume a simple power law 
\begin{equation} \label{eqn:lhcp}
	\lambda^h_c=c_1 (\lambda^h_{max})^{c_2}.	\end{equation}
On the other hand, since by Eqn.~\eqref{eqn:lnch} the permanent and contour stretches are related by 
	$ \lambda^h_n=\left( \lambda^h_c / \lambda^h_{co} \right)^\frac{1}{2}$,
where $\lambda^h_{co}\equiv c_1$ is the initial contour stretch, by Eqn.~\eqref{eqn:lhcp} the natural stretch is given by\begin{equation} \label{eqn:lncp}
	\lambda^h_n
	=(\lambda^h_{max})^{\frac{c_2}{2}}.	\end{equation}
We point out that according with previous microstructure based analysis the natural and the contour stretches are analytically related (here assigned by the two only constants $c_1$ and $c_2$).
If then, as anticipated, we consider a damage  taking care of the described humidity induced crystal misalignment, the stress-stretch relation for the hard part by using Eqns.~(\ref{eqn:force-stretch}) and (\ref{eqn:stretch measures})  is
\begin{equation} 	\label{eqn:sigma stretch hard part}
	\sigma^h = 
	\hat E^h(\text{\small RH}) \left[ 
	\left( \frac{\hat \lambda^h_c(\lambda^h_{max})-\hat\lambda^h_n( \lambda^h_{max})}{\hat\lambda^h_c( \lambda^h_{max})-\lambda^h}\right)^2-1
	\right] \end{equation}
with
\begin{equation}	\label{eqn:ehb}
	E_h=\hat E^h(\text{\small RH})=(1-\alpha \ \text{\small RH})E^h_o,	\end{equation}
where $E^h_o=N^h_{fib} \kappa^h$ is the modulus in the dry condition  and $\alpha$ measures the humidity induced damage rate.

Eventually, by following \cite{flory_1982}, we consider a matrix embedding hard and soft fractions describing the complex macromolecular network composing the spider thread, with inner and intrachains connections. The total free energy is then calculated as the sum of the energy of ideally isolated hard and soft chains described above and elongated along the fiber direction plus an energy term taking care of chains interactions.
Under a simple Neo-Hookean assumption (corresponding at the molecular scale to harmonic network chains \cite{rubinstein2003polymer}) for incompressible material, the matrix stress for a uniaxial extension is		\begin{equation}
		\label{eqn:sigma stretch matrix}
		\sigma^m = \mu\left(\lambda^h -\frac{1}{(\lambda^h)^2} \right), 
	\end{equation}
	with $\mu$ the material shear modulus. Observe that we assume that the matrix natural length  coincides with the hard fraction initial one, so that $\lambda^m=\lambda^h$.

\begin{figure}[h!] \vspace{-0.0 cm}
	\centering \includegraphics[width=.5\textwidth]{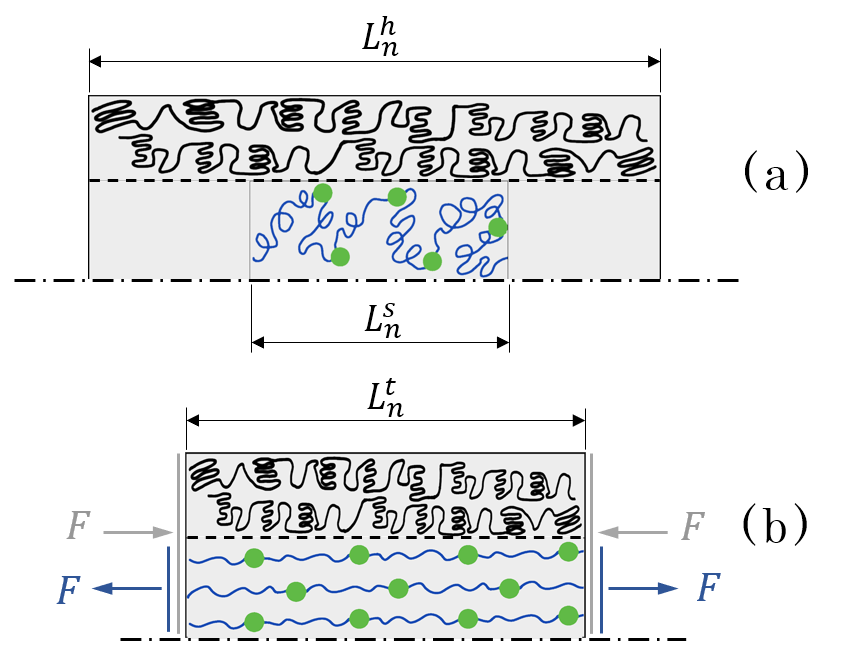} \vspace{-0.05 cm}
	\caption{\label{fig:prestretch}
		Outer and inner regions have different natural lengths (a) leading to a prestretch of the internal amorphous part (b).} \vspace{-0.4cm} \end{figure}

We are now in the position of deducing the overall behavior of the spider thread. Our deduction is based on the main assumption that the spun initial length of the fractions are the same: $n^h b^h = n^s b^s$ (kinematic compatibility). After exposition to humidity the soft region chains reduce their natural length (see Fig. \ref{fig:prestretch}).
The kinematic compatibility impose that the different regions undergo the same stretch $\lambda^t_o$ that corresponds to zero overall initial stress. On the other hand, when the fiber is subjected to a force $F>0$, it undergoes a stretch $\lambda^t$, starting from the new natural configuration $\lambda^t_o$. As a final result the stretches for the soft region (that varies its natural stretch according to humidity), hard region and matrix starting from the spun initial length (see the scheme in Fig.~A5 of the Appendix) are given by 
$\lambda^s=	\lambda^h=
\lambda^m=\lambda^t \lambda^t_o$,
 where $\lambda^t$ represents the experimentally measured stretch. 

The overall stress-stretch relation is	$$
\sigma^t(\lambda^t)=\Theta (\lambda^h_n-\lambda^t \lambda^t_o) (1-\alpha \ \text{\small RH})E^h_o\left[ 
\left( \frac{\lambda^h_c-\lambda^h_n}{\lambda^h_c-\lambda^t \lambda^t_o} \right)^2-1 \right] \vspace{- 0.5cm} $$
\begin{equation}\label{eqn:total stress ext}  +E^s\left[ 
\left( \frac{\lambda^s_c-\lambda^s_n}{\lambda^s_c-\lambda^t \lambda^t_o} \right)^2-1 \right]
+\mu\left(\lambda^t \lambda^t_o-\frac{1}{( \lambda^t \lambda^t_o)^2} \right)	\end{equation}
where $\Theta$ is the step function considering that the hard fraction chains are not able to sustain any compressive force ($\sigma^h =0$ if $\lambda^h<\lambda^h_n$). Observe that from this equation at $\lambda^t=1$ and $\sigma^t=0$ we determine  $\lambda^t_o$.

A comment about a second important aspect of the variable mechanical behavior is now in order: {\it temperature effects}. Indeed, when the temperature at fixed {\small RH} is increased, the silk undergoes an effect of link scissions as described for humidity~\cite{plaza_thermo-hygro-mechanical_2006}. Moreover, also temperature growth is accompanied by fiber contraction \cite{glisovic_temperature_2007} and again the experiments show the existence of a critical value
where such effects of link scission and length variation are strongly localized. In analogy with polymer mechanics this value is indicated as glass transition temperature~$T_g$.
In particular, in \cite{fu_moisture_2009} the authors obtained an experimental linear relation between $T_g$ and {\small RH}. Of course such a relation would ask a theoretical description that by itself appears to be very interesting, but it is out of the aims of this paper. Instead, to show that our model can reproduce also the experimental temperature effects, we {\it phenomenologically} assume a Gaussian dependence of the number of links from temperature in Eqn.~\eqref{eqn:drh} (where {\small RH} is substituted by $T$) and then we modify correspondingly  the constitutive equation Eqn.~\eqref{eqn:total stress ext}. Accordingly {\small RH}$_c$ is substituted by   $T_g$. The efficacy of these assumptions are again fully supported in the following section.

The final aspect of the model regards the humidity and temperature dependence of the limit stretch: to this hand we need a {\it fracture criterion}. 
Based on the considerations in \cite{yazawa_simultaneous_2020} 
we here assume that the fracture is regulated by the hard fraction and in particular that the fracture condition is $\lambda^h=\lambda_{lim}$, where $\lambda_{lim}$ is a given constitutive parameter. As we show in the following section, this criterion is successful with the exception of the fully dry condition where the breakage is known to be induced by localized damage defects \cite{yazawa_simultaneous_2020}.\vspace{0.2 cm}

\section{Experimental validation}
 
\begin{figure}[htb]	
	\includegraphics[width=.52\textwidth]{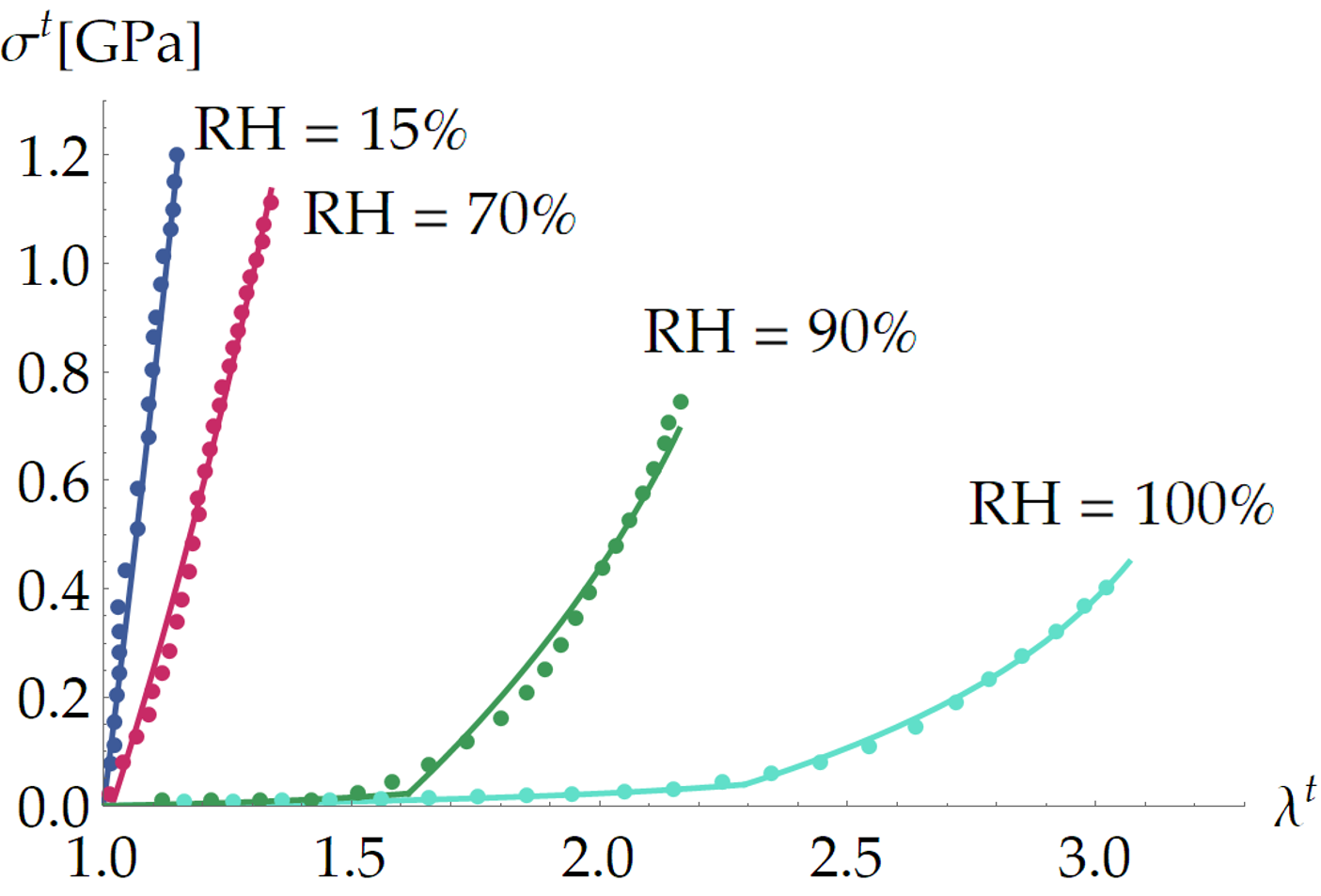}\vspace{-0.2 cm}
	\caption{\label{fig:exp2} Theoretical (continuous lines) {\it vs} experimental (dots, reproduced from \cite{elices_finding_2005}) stress-stretch curves for  {\it Argiope trifasciata} spider fibers at different {\small RH} and $T=20\,^\circ \!$C. Here  $k^h=2.2 $ GPa, $k^s=13.5 $~MPa, $\mu=0.14$~MPa, $c_1=1.3,\ c_2=0.87,$ $\alpha=0.0094,\ s=8.5,\ m_f/m_o=0.12,\ \lambda^s_c=1.62,\ \lambda_{lim}=1.34$.
	}\vspace{-0.2 cm} \end{figure}

\noindent In this final section we verify the effectiveness of the proposed model by quantitatively comparing the main experimental effects induced by humidity and temperature variations on different spider silks with the theoretical behavior. Consider first the tensile response under variable $\text{\small RH}$ for a highly stretchable silk (Argiope trifasciata fibers, reproduced by~\cite{elices_finding_2005}). As shown in Fig.~\ref{fig:exp2} this silk exhibits a remarkable dependence of the mechanical response on humidity.
We may observe two different regimes (see details in Fig. A6 in the Appendix) in accordance with the silk experimental response: for {\small RH}$<${\small RH}$_c$ the behavior is almost linear and this is due in our model to the dominance of the hard fraction; for {\small RH}$>${\small RH}$_c$ we have two regimes. Initially the silk is highly stretchable, with high deformations at very low forces. In this regime the numerical simulations show that the hard region is shorter than its natural length, so it does not contribute to the fiber stiffness. When this length is attained, the fiber exhibits a sudden hardening.
In Fig.~\ref{fig:exp2} it is possible also to verify the efficacy of the introduced fracture criterion. Indeed, we calibrated the hard fraction limit stretch to reproduce the experimental limit at {\small RH} = 70\% and then we predicted the  {\small RH} = 90\% and 100\% cases with errors of only $0.17\%$ and $1.49\%$, respectively.
As anticipated  the prediction is less accurate for very low humidities.	

\begin{figure}[t]
	\centering 
	\includegraphics[width=.7\textwidth]{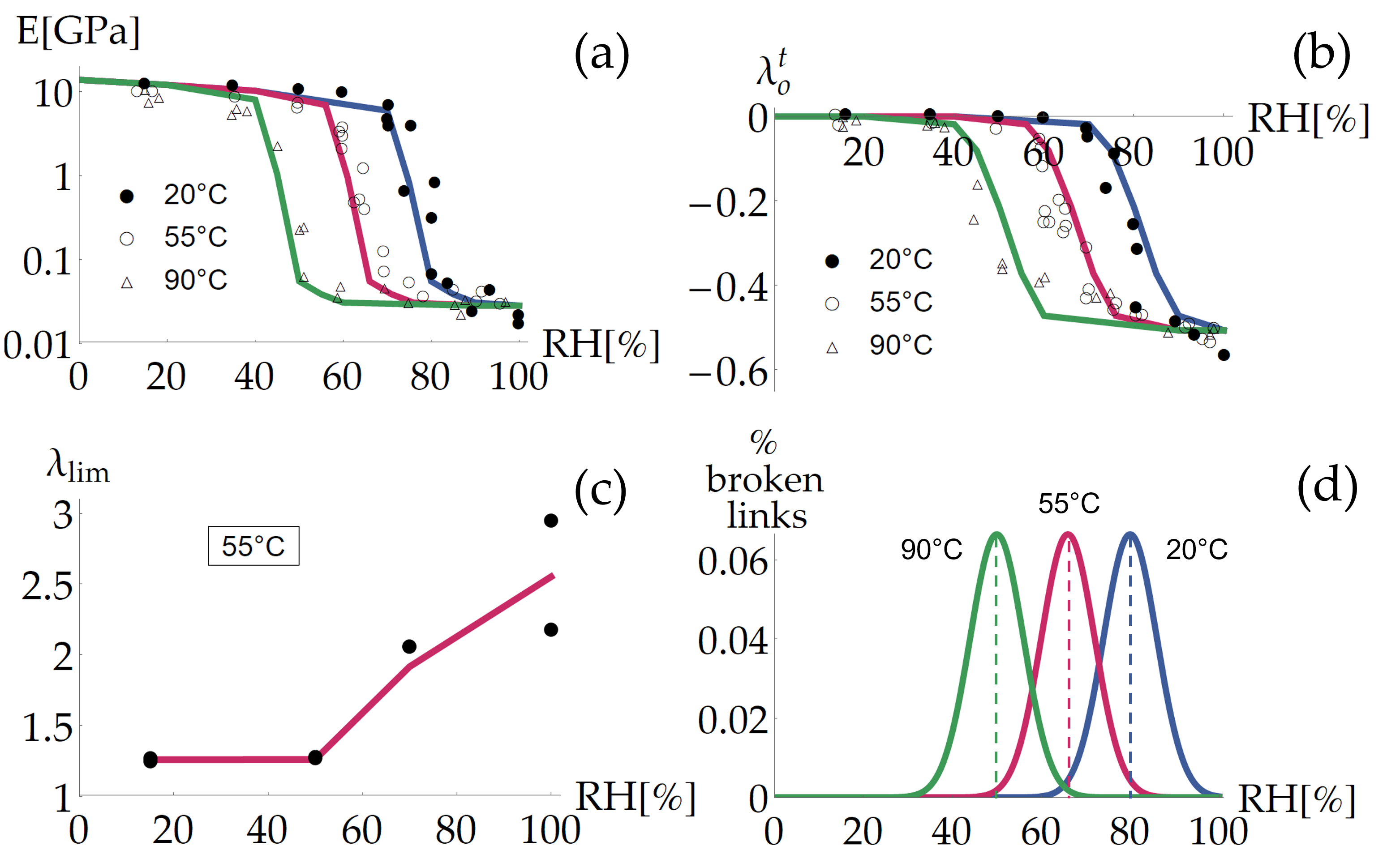}\vspace{-0.3 cm}
	\caption{\label{fig:elo} Theoretical (continuous lines) {\it vs} experimental (dots) curves for the initial elastic modulus (a), initial (zero force) stretch (b), limit stretch (c) and assumed Gaussian distribution of broken links (d) as a function of {\small RH} for \textit{Argiope trifasciata} fibers \cite{plaza_thermo-hygro-mechanical_2006}. Here $k^h=4.05$ GPa, $
	k^s=39.6 $~MPa, $\mu=0.2$~MPa, $c_1=1.4,\ c_2=0.75,\ \alpha=0.0065,$ $s=0.6,$ $m_f/m_o=0.23,$ $\lambda^s_c=2.05, \ \lambda_{lim}=1.26,$ and {\small RH}$_c=80,66,50\%$ for $T=20,55,90\,^\circ \!$C respectively.\vspace{-0.5 cm}
	}\vspace{-0.0 cm} \end{figure}

To further test the efficiency of the model, we show the possibility of predicting the influence of humidity and temperature on other important material parameters, such as elastic modulus, supercontraction stretch of unrestrained fibers and limit stretches (Fig.~\ref{fig:elo}).
 In the prediction of the experiments we fixed the material parameters and changed only {\small RH}$_c$ at different temperatures using the experimental values in~\cite{plaza_thermo-hygro-mechanical_2006}.
We remark that the limit stretch is reproduced for the only available testing temperature ($55\,^\circ \!$C) (see also details in the Appendix).  

\begin{figure}[htb]\vspace{-0.2 cm}
\includegraphics[width=.45\textwidth]{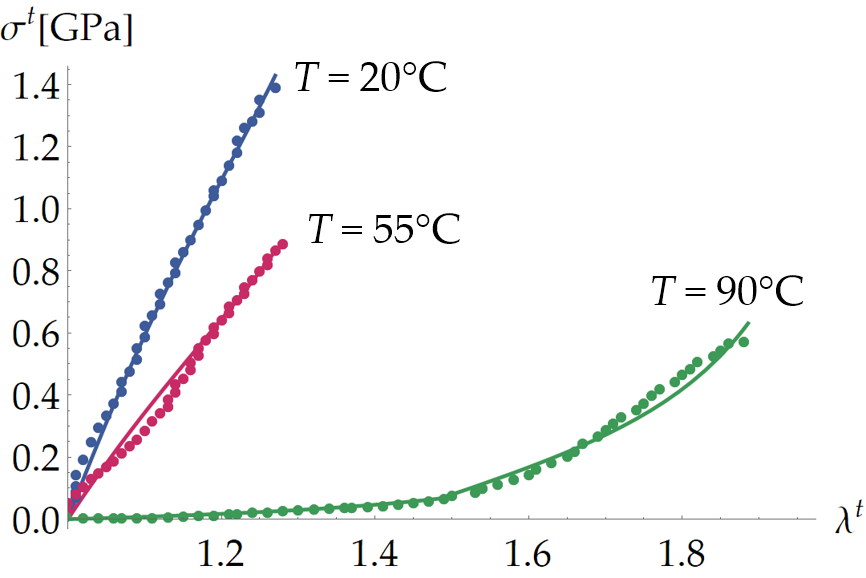}\vspace{-0.3 cm}
	\caption{\label{fig:temp} Theoretical (continuous lines) {\it vs} experimental (dots) stress-stretch curves for {\it Argiope trifasciata} spider fibers at different  temperature at fixed {\small RH} $=50\%$ (reproduced from \cite{plaza_thermo-hygro-mechanical_2006}).
		Here $T_g=84\,^\circ \!$C$,\ k^h=3.83 $~GPa, $k^s=32.7 $~MPa, $\mu=2$~MPa, $c_1=1.36,$ $c_2=1.25,\ \alpha=0.00995,$ $s=4.5,$ $m_f/m_o=0.345,\ \lambda^s_c=1.49,\ \lambda_{lim}=1.27$.
	} \vspace{-0.1 cm} \end{figure}

As a final experimental comparison we consider the effects induced by variable temperature at fixed {\small RH} in~\cite{plaza_thermo-hygro-mechanical_2006}. We evaluated $T_g$ at given {\small RH}=50$\%$  using the relation reported in the same paper. The results exhibited in Fig.~\ref{fig:temp} show again an accurate reproduction of the experiments.\vspace{0.15 cm}

To conclude, we remark that the described ability of the proposed model of predicting the experimental behavior of different mechanical properties make us confident that it well reproduces the described humidity and temperature effects at the molecular scales. 
This is supported even more in the  Appendix by predicting the behavior of different silks and environmental conditions. We strongly believe that this is a consequence of our microstructure deduction of the material response function.
The physical meaning of all the adopted (microscopic) parameters opens up the possibility of applying the proposed model not only to other  protein materials with similar structures \cite{puglisi_micromechanical_2017}, but also in the design of bioinspired materials employing chosen specific proteins \cite{Greco2021,liu_spider_2019}.

\newpage Funding: GP has been supported by the Italian Ministry MIUR-PRIN project 2017KL4EF3 and by GNFM (INdAM), DD  by the Italian Ministry MIUR-PRIN project 2017J4EAYB and NMP by the European Commission under the FET Open “Boheme” grant no. 863179 and by the Italian Ministry of Education MIUR under the PON ARS01-01384- PROSCAN Grant and the PRIN-20177TTP3S.

Acknowledgments: The authors thank G. Greco for insightful discussions on the subject of the paper and for indicating important papers missing in the previous version including the very recent article \cite{cohen_origin_2021} that suggests the role of the humidity induced variation of the molecules natural configuration. \vspace{- 0.5 cm}

\bibliography{PRL}% Produces the bibliography via BibTeX.

\newpage

\section{Appendix}

\renewcommand{\theequation}{A\arabic{equation}}
\renewcommand{\thefigure}{A\arabic{figure}}

\setcounter{equation}{0}
\setcounter{figure}{0}

\noindent We here detail some physical and analytical properties  of the different regions. Moreover in the successive section we propose further experimental comparisons. 

\subsection{Model Equations}
\noindent Here we give some details on the analytical deduction of the constitutive model starting from the described microstructure interpretations. \vspace{0.3 cm}

\noindent {\bf Soft region.} The contour and natural lengths of this fraction varies with {\small RH} and can be calculated, for ideal chains, as $ L^s_c=n^s \ b^s, L^s_n=\sqrt{n^s} \ b^s$
where $n^s$ is the (mean) number of Kuhn segments with length $b^s$. We identify the number of H-bonds with the number of domains in which the chain is divided (see Fig.~1(b)).
Thus, starting from its initial value $n^s_o$, the number of monomers depends on the number of H-bonds $n^s=\frac{n_o^s}{m}$.

Accordingly, the natural length of the amorphous part can be expressed as 
\begin{equation} L^s_n=m\sqrt{\frac{n_o^s}{m}} \ b^s=\sqrt{n_o^s \ m} \ b^s \end{equation}
so that, the initial natural length is 
$L^s_o=\sqrt{n_o^s m_o} \ b^s$.
The natural stretch can be therefore written as a function of the number of H-bonds as
\begin{equation} \label{eqn:lanb}	\lambda^s_n=\frac{L^s_n}{L^s_o}=
	\frac{\sqrt{n_o^s \ m} \ b^s}{\sqrt{n_o^s m_o} \ b^s}=\sqrt{\frac{m}{m_o}}. \end{equation}
	
If we consider a Gaussian distribution for the breaking rate of the bonds	
as in Eqn.~2 we get the evolution of the natural length of the soft fraction as
\begin{equation} \lambda^s_n=\sqrt{1+d(\text{\small RH})\left( \frac{m_f}{m_o}-1 \right)}. \end{equation}

\begin{figure}[t!]
	\includegraphics[width=.62\textwidth]{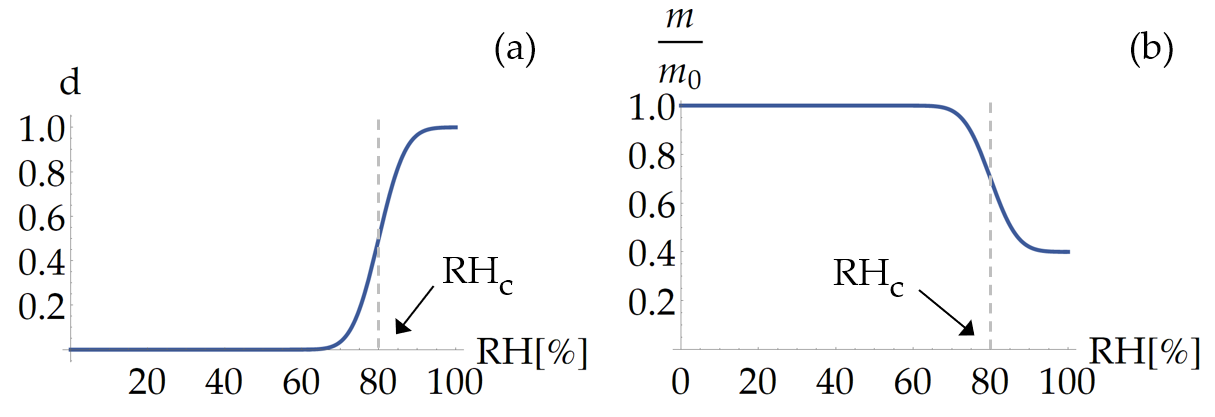}\vspace{-0.3cm}
	\centering
	\caption{\label{fig:mrh} 
		(a) Damage parameter $d$ representing the percentage of broken links as a function of humidity.  
		(b) Influence of the relative humidity on the number of H-bonds with respect to the initial number of H-bond. 
		Here $\text{\small RH}_c=80\%$, $s=5.5$ and $m_f/m_o=0.4$.}
\end{figure}

\begin{figure}[b]
	\centering
	\includegraphics[width=.45
	\textwidth]{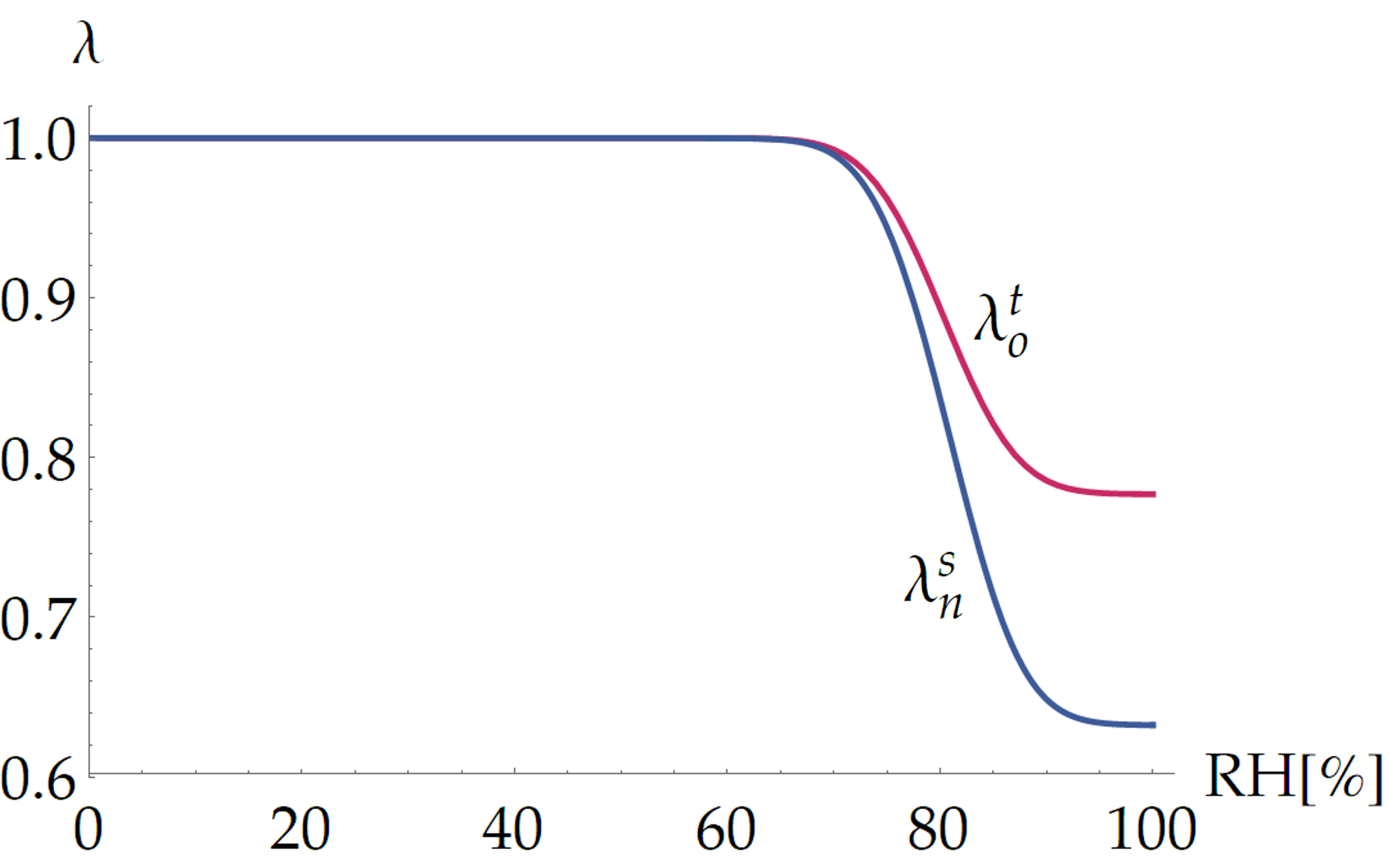}
	\caption{\label{fig:trends} Dependence of the natural configuration (natural stretch, in blue) of the soft region $\lambda^s_n$ from the relative humidity. Observe the abrupt decrease around the supercontraction threshold $\text{\small RH}_c=80\%$.
	Here $s=5.5$ and  $m_f/m_o=0.4$. For comparison, the unstressed stretch of the overall fiber $\lambda^t_o$ is represented in pink. It is calculated assuming $k^s=1.35$ MPa$,
		\mu=2.5$ MPa$,
		\lambda^s_c=2.62$.
	}
\end{figure}
A typical variation of the damage function and number of domains  under our Gaussian probability choice is represented in Fig.~\ref{fig:mrh}, whereas the variation of the natural stretch of the soft fraction is reported in Fig.~\ref{fig:trends}.
We remark that the parameters needed to compute the variation of the natural stretch as a function of humidity are $m_f/m_o$, \text{\small RH}$_c$ and $s$.
On the other hand, the corresponding expression for the contour length is
$L^s_c={m\frac{n_o^s}{m}} \ b^s =n_o^s b^s,$
so that the contour stretch of the amorphous part results constant
$ \lambda^s_c=\frac{L^s_c}{L^s_o}=
	\frac{n_o^s \ b^s}{\sqrt{n_o^s \ m_o} \ b^s}=
	\sqrt{\frac{n_o^s}{m_o}}. $
	 \vspace{0.3 cm}

\begin{figure}[ht]
	\centering
	\includegraphics[width=.5\textwidth]{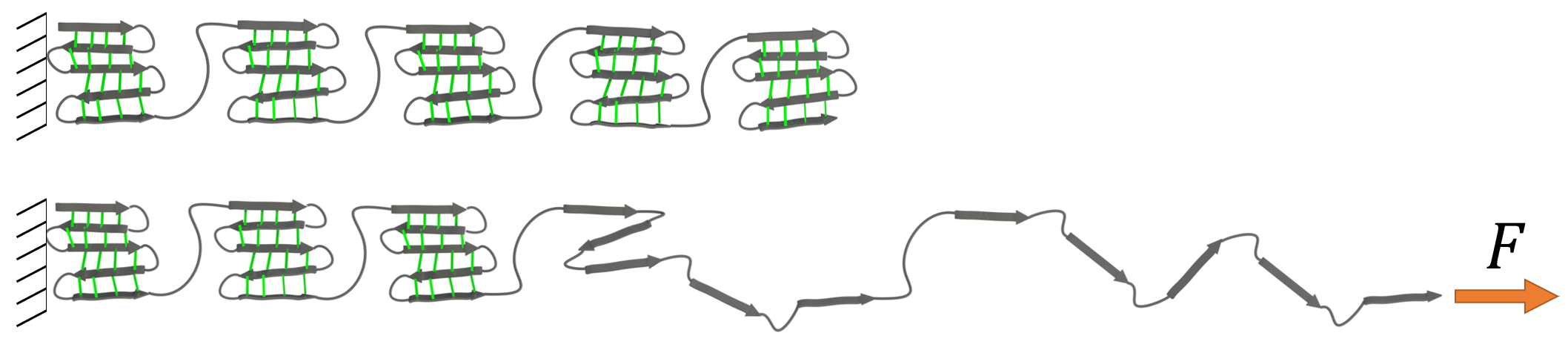}\vspace{-0.2cm}
	\caption{\label{fig:h_unfold}
		Cartoon of a chain of the hard region undergoing a folded $\rightarrow$ unfolded transition when subjected to a force.}
\end{figure}

\noindent {\bf Hard region.} This fraction, characterized by a high percentage of nano crystallites typically in the $\beta$-sheet conformation, can undergo a phenomenon of crystal domains unfolding induced by chains stretching (see the scheme in Fig.~\ref{fig:h_unfold}). 

\noindent Thus, we assume that the number of monomers of a macromolecule depends on the maximum length attained by this fraction $n^h=\hat{n}^h(L^h_{max})$. Correspondingly, the contour and natural lengths, for ideal chains, are respectively given by $L^h_c=n^h \ b^h, L^h_n=\sqrt{n^h} \ b^h$
with $b^h$ the Kuhn length of a macromolecule in the hard region. 
Then, let $n^h_o$ be the initial number of monomers, the initial natural length
$L^h_o=\sqrt{n^h_o} \ b^h $
 can be used to calculate the contour stretch as
\begin{equation}
	\label{eqn:lhc}
	\lambda^h_c=\frac{L^h_c}{L^h_o}
	=\frac{n^h b^h}{\sqrt{n^h_o}\, b^h}
	=\frac{n^h}{\sqrt{n^h_o}}
	=\frac{\hat n^h (\lambda_{max})}{\sqrt{n^h_o}}
\end{equation}
and the natural stretch as
\begin{equation}
	\label{eqn:lhn}
	\lambda^h_n=\frac{L^h_n}{L^h_o}
	=\frac{\sqrt{n^h}\, b^h}{\sqrt{n^h_o}\, b^h}
	=\sqrt{\frac{n^h}{n^h_o}}
	=\sqrt{\frac{\hat n^h (\lambda_{max})}{n^h_o}}.
\end{equation}

\begin{figure}[htb]
	\centering
	\includegraphics[width=.5\textwidth]{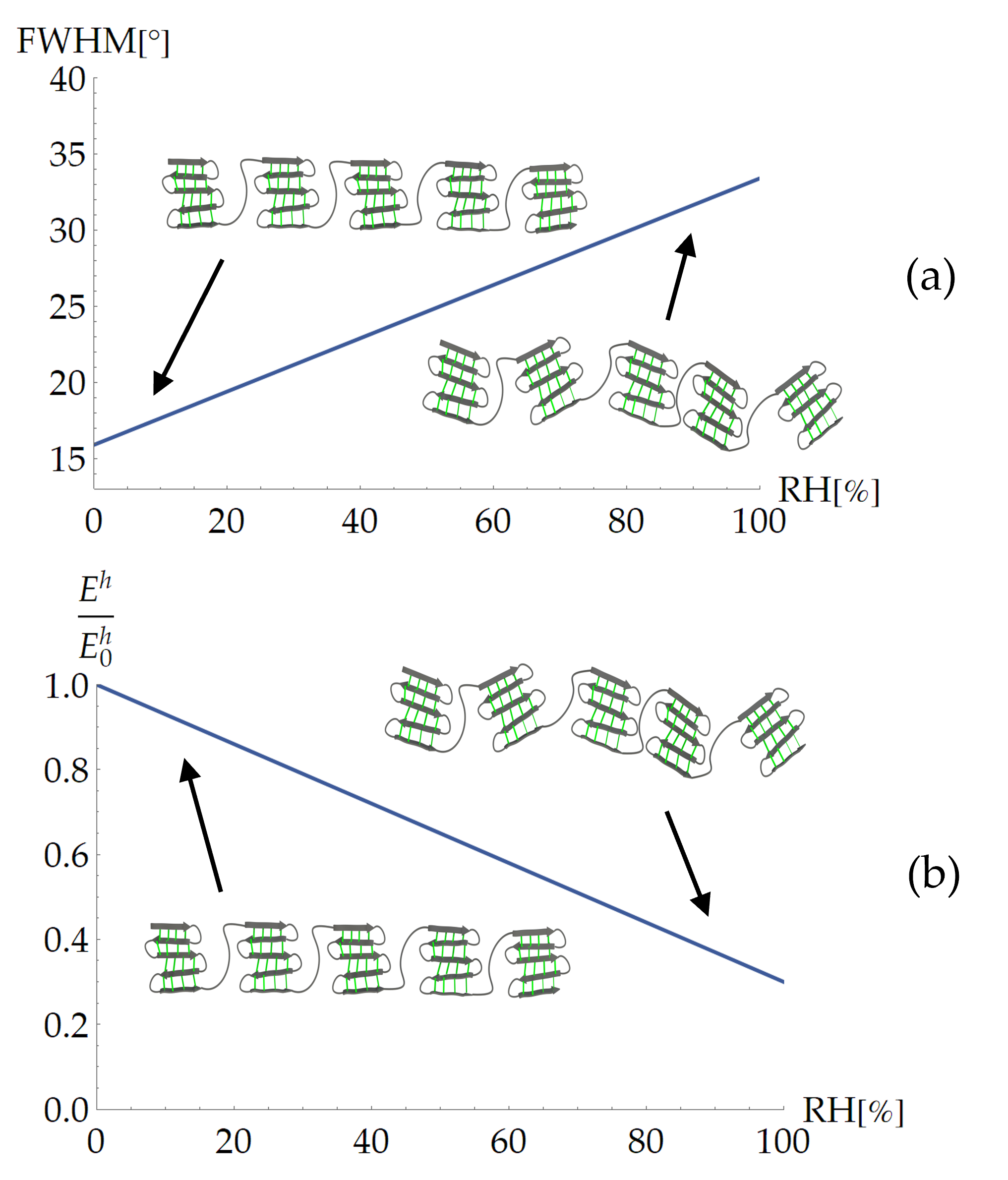} \vspace{-0.2cm}
	\caption{\label{fig:h_humid}
		(a) WAXS measurements of dragline silk fibers  at different {\small RH}s are used to calculate the  orientation of crystalline $\beta$-sheets with respect to the fiber axis \cite{yazawa_simultaneous_2020} with the result that FWHM linearly increases with {\small RH}. This proves a reduced orientation of the nano crystals under higher {\small RH} conditions as schematized by the cartoons.
		(b) Assumed damage function depending on the relative humidity. The reduction of the elastic modulus of the hard part is associated to the lack of orientation of the crystal as the humidity increases. Here we assume $\alpha=0.007$.}\vspace{-0.3cm}
\end{figure}

The second important variation induced in the crystal domains is related to humidity that reduces the orientation of the nanocrystallites, as reported in Wide-Angle X-ray Scattering (WAXS) measurements in Fig.~\ref{fig:h_humid}(a) reproduced from \cite{yazawa_simultaneous_2020}. 
In particular, the Full Width at Half Maximum (FWHM) value increases linearly with $\text{\small RH}$, thus indicating a lower orientation of the nano crystals at higher $\text{\small RH}$ values.
Moreover, the crystallites misalignment affects the elastic modulus, as experimentally verified in \cite{du_design_2006}.
We therefore introduce the humidity effect in the crystalline region by considering a damage function linearly dependent from $\text{\small RH}$ (Eqn.~12). The resulting modulus variation is represented in Fig.~\ref{fig:h_humid}(b).

%Finally the stress-stretch equation of the hard part is
%\begin{equation}\label{eqn:sigma stretch soft part} \sigma^h = E^h(\text{\small RH})\left[ \left( \frac{\lambda^h_c-\lambda^h_n}{\lambda^h_c-\lambda^h}\right)^2-1\right].\end{equation}
%with the contour and the natural stretch given by \eqref{eqn:lhcp} and \eqref{eqn:lncp} respectively.

 \vspace{0.3 cm}

\begin{figure*}[htb]	
	\centering
	\includegraphics[width=.95\textwidth]{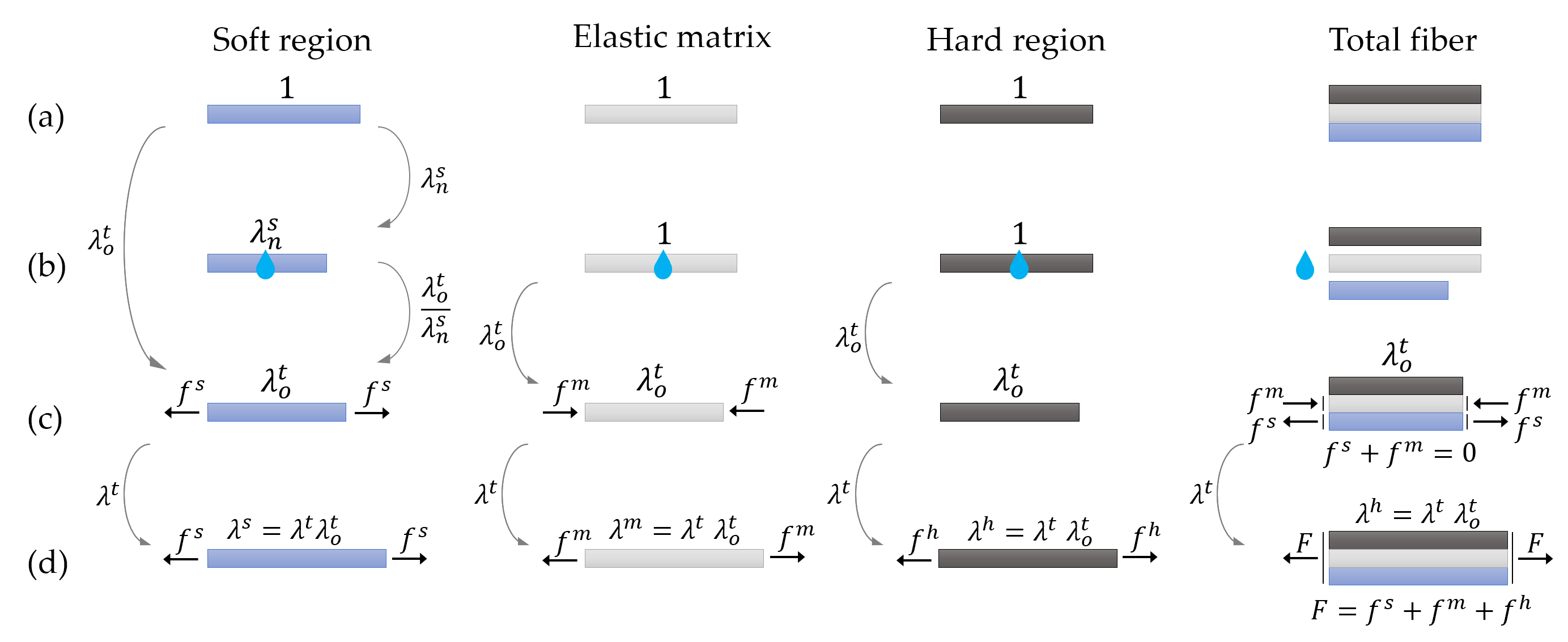}\vspace{-0.4 cm}
	\caption{\label{fig:matching}Scheme of the evolution of the configurations of the different composing phases and of the whole fiber: (a) dry natural configuration, (b) humid condition without external forces where selfstresses (reported in (c)) are induced by kinematic compatibility, (d) non zero external force configurations.}
\end{figure*}

\noindent {\bf Overall fiber.}
The evolution of the natural length of the different fractions and the insurgence of self-stretches and internal forces under different humidity and force conditions are schematized in Fig.~\ref{fig:matching}.

\begin{figure*}[htb] \centering
	\includegraphics[width=1\textwidth]{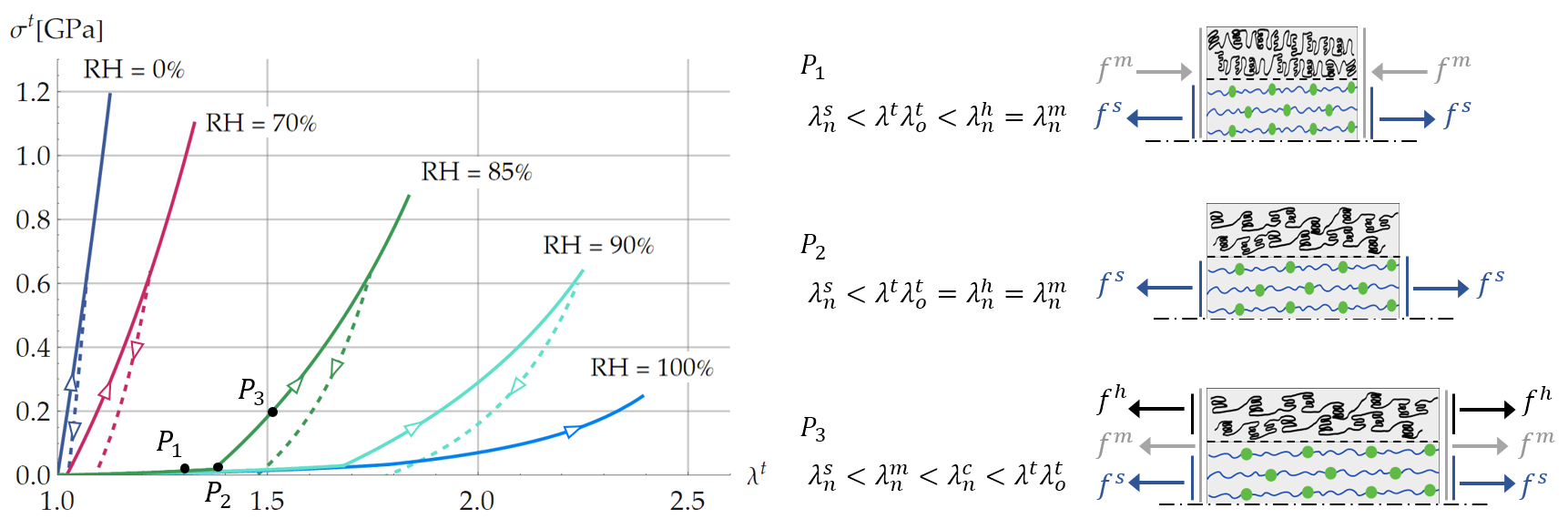}\vspace{-0.4 cm}
	\caption{\label{fig:model} Theoretical stress-stretch curves at different humidity conditions (continuous lines correspond to loading and dashed to unloading curves). Three points of the curve {\small RH} $=85\%$ are marked to illustrate  different regimes schematically illustrated on the right: $P_1$ hard phase is unloaded, $P_2$ hard phase in its natural configuration, $P_3$ hard phase in traction. Here $k^h=2.16$~GPa, $k^s=18$~MPa, $\mu=0.14$~MPa, $c_1=1.33,\ c_2=0.75,\ \alpha=0.0099,\ m_f/m_o=0.3,\ s=6.5,\ \lambda^s_c=1.65, \text{{\small RH}}_c=80\%, \lambda_{lim}=1.34	$.} \vspace{-0.2cm} 
\end{figure*}

The overall stress-stretch relation as a function of the experimental stretch $\lambda^t$ can be deduced by Eqn.~14.
In Fig.~\ref{fig:model} we describe the behavior of the model here proposed, representing the stress-stretch curves for different humidity conditions ({\small RH} $=0\%,70\%,85\%,90\%,100\%$). 

Starting from the dry condition ($\text{\small RH}=0\%$), here the natural length of the hard and soft fraction coincide ($\lambda_n^h=\lambda_n^s$). As a result the hard fraction participates to the mechanical response from the beginning. Consequently the force is carried mainly by the much stiffer hard fraction in all the elongation experiment. The behavior is similar for $\text{\small RH}<\text{\small RH}_c$ where a lower stiffness is due to the hard domain disorder inducing a damage according with Fig.~\ref{fig:h_humid} and Eqn.~12 (see the curve  $\text{\small RH}=70\%$).
Once $\text{\small RH}>\text{\small RH}_c$, the role played by the amorphous fraction becomes more relevant because its natural length $\lambda^s_n$ undergoes a significant decrease (see Fig.~\ref{fig:trends}). This leads to a consequent substantial decrease of the initial (zero force) length of the total fiber $\lambda^t_o$. Thus, as long as the fibrils of the hard region do not reach again their natural length, the mechanical response is given only by the amorphous phase and the matrix (see {\it e.g.} point $P_1$ in Fig.~\ref{fig:model}). 
Then, as soon as the hard region starts to be stretched (point $P_2$) the force starts again to be mainly sustained by the stiffer crystalline phase (see {\it e.g.} point $P_3$). This transition is evidenced both in the theory and in the experiment by a `localized' material hardening.
The behavior is similar also for higher values of the humidity  ($\text{\small RH}=90\%$), with different values of $\lambda_t$ distinguishing the two regimes. 
At extreme humidity conditions (last curve at  $\text{\small RH}=100\%$) the mechanical behavior may be given by the only amorphous phase and matrix if the ultimate stretch of the fiber is lower than the transition threshold. 
Even though in this paper we  focussed on the humidity effects on the monotonic stress-stretch curves, in Fig.~\ref{fig:model} we also show through dashed lines the system behavior when subjected to unloading. This figure let us show that, based on the microstructure interpretation, the proposed model is able to describe not only the fundamental macroscopic damage effect, but also the experimentally observed presence of residual stretches \cite{vehoff_mechanical_2007}. Interestingly, permanent deformations are not deduced as usually independently from damage, {\it e.g.} through the introduction of new  variables, whereas both damage and residual stretches descend from the unfolding of the hard domains. 

 \subsection{Experimental validation}
\begin{figure}[htb]
	\centering \vspace{-0.1 cm}
	\includegraphics[width=.47\textwidth]{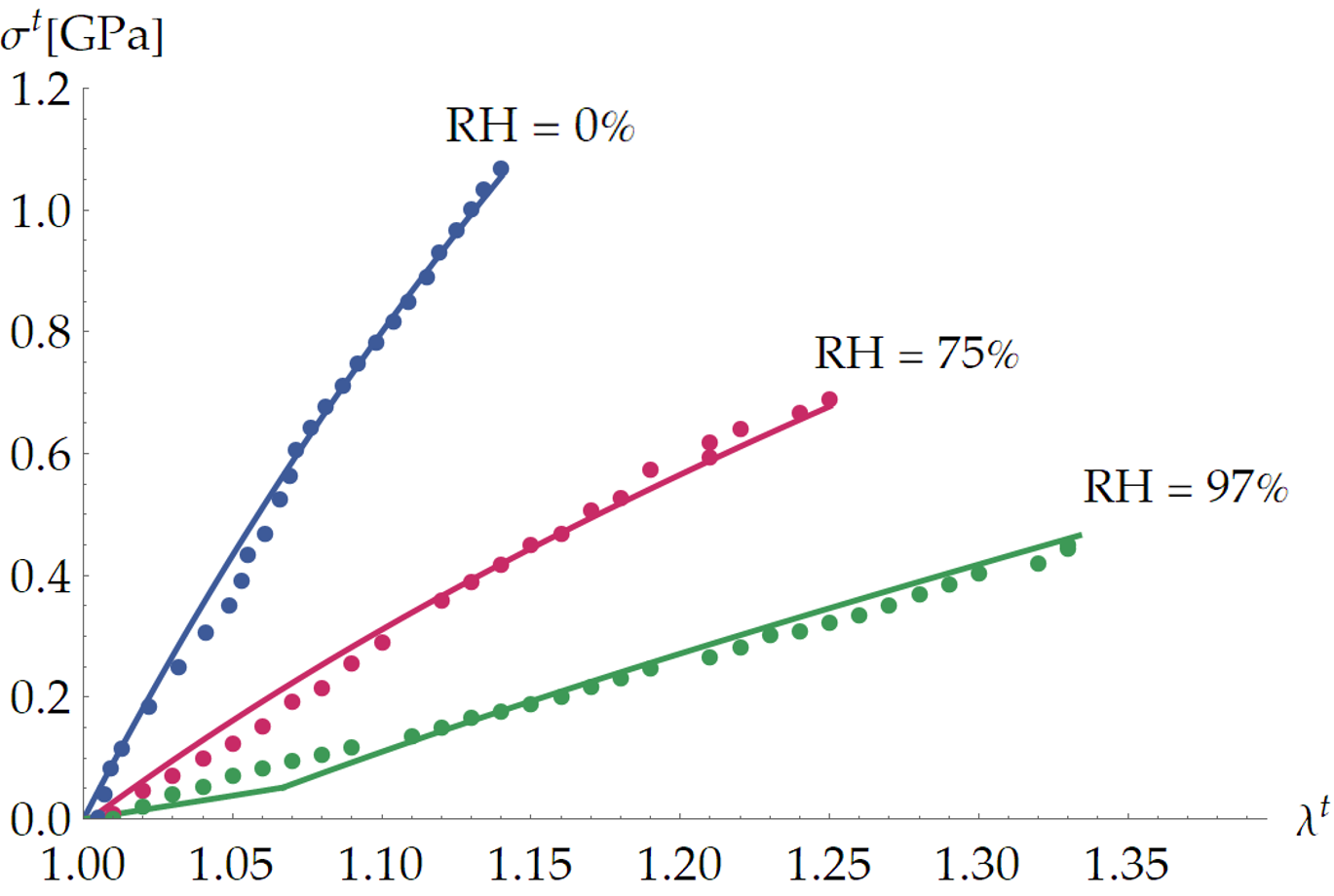} %\vspace{-0.1 cm}
	\caption{\label{fig:exp1} Theoretical (continuous lines) {\it vs} experimental (dots) stress-stretch curves for \textit{Nephila clavata} fibers at different humidity and $T=25\,^\circ \!$C  (reproduced from the experiments at  strain rate of $3.3 \times 10^{-3} \ s^{-1}$ in \cite{yazawa_simultaneous_2020}). Here $k^h=5.6$~GPa, $k^s=1.47$~GPa, $\mu=1.1$~MPa, $c_1=1.35,\ c_2=1.46,\ \alpha=0.009,$  $m_f/m_o=0.878,\ s=3,\ \lambda^s_c=4.6, \lambda_{lim}=1.25, \  \text{\small RH}_c=80\%.$}
\end{figure}

\noindent To test the effectiveness of the proposed model in describing quantitatively the experimental behavior, 
in addition to the {\it Argiope trifasciata} spider fibers (Fig.~3), in this SM we consider tensile tests performed on a \textit{Nephila clavata} spider fiber under various $\text{\small RH}$s ($0\%, 75\%, 97\%$) reproduced from \cite{yazawa_simultaneous_2020}.
In Fig.~\ref{fig:exp1} we report the comparison between the experimental results and the theoretical model. Despite this silk shows a remarkably different response to the humidity variations, the proposed model is significantly successful in predicting the observed experimental behavior.
In Fig.~\ref{fig:exp1} we also test the above described rupture hypothesis for the \textit{Nephila clavata} fibers by using the value corresponding to the experimental break at {\small RH} $=75\%$ to predict the breaking strain at  {\small RH} $=97\%$ with an error of $0.22\%$. This remarkable small error confirms the plausibility of the proposed rupture criterion.
Observe anyway that this hypothesis in this silk cannot be applied to the fully dry case where the breakage is typically induced by localized defects \cite{yazawa_simultaneous_2020}.

\begin{figure}[htb]
	\centering
	\includegraphics[width=.5\textwidth]{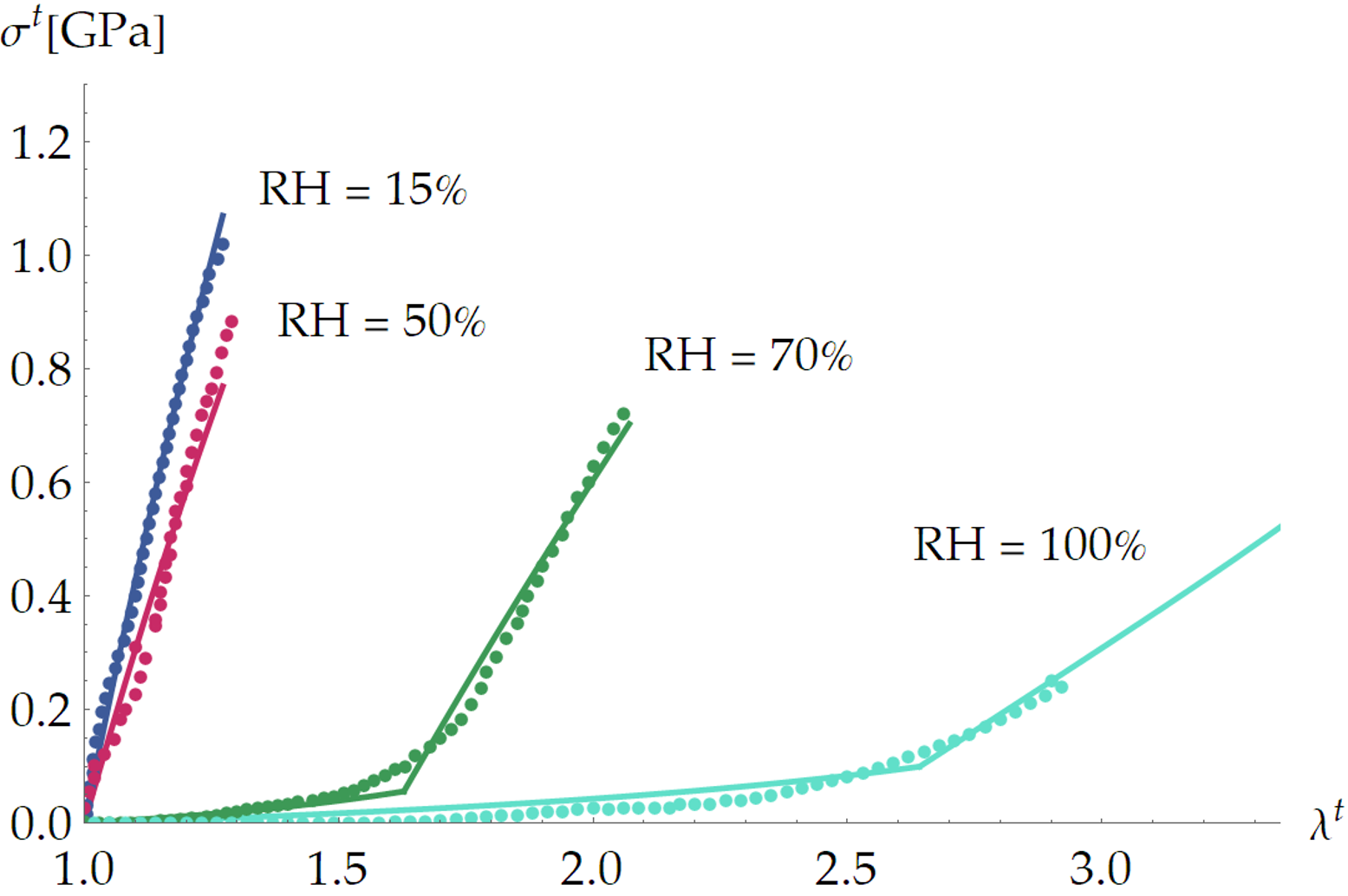} \vspace{-0.15 cm}
	\caption{\label{fig:exp3} Theoretical (continuous lines) {\it vs} experimental (dots) stress-stretch curves for {\it Argiope trifasciata} spider fibers at different  humidity and $T=55\,^\circ \!$C (reproduced from \cite{plaza_thermo-hygro-mechanical_2006}). Here  $k^h=0.99 $~GPa, $k^s=18.2 $~MPa, $\mu=0.01$~MPa, $c_1=1.18,\ c_2=1.1,$$\alpha=0.0085,\ m_f/m_o=0.162,\ s=5.5,\ \lambda^s_c=1.57,\ \lambda_{lim}=1.27, \ \text{\small RH}_c=66.5\%.$}
\end{figure}

In Fig.~\ref{fig:exp3} we show the ability to reproduce the stress-stretch behavior reported in \cite{plaza_thermo-hygro-mechanical_2006} also at high testing temperature ($T=55\,^\circ \!$C). The model turns out to be successful in the prediction of the humidity effects at high Temperatures in the stress-stretch test.

\begin{figure}[h]
	\centering \vspace{0.21 cm}
	\includegraphics[width=.4\textwidth]{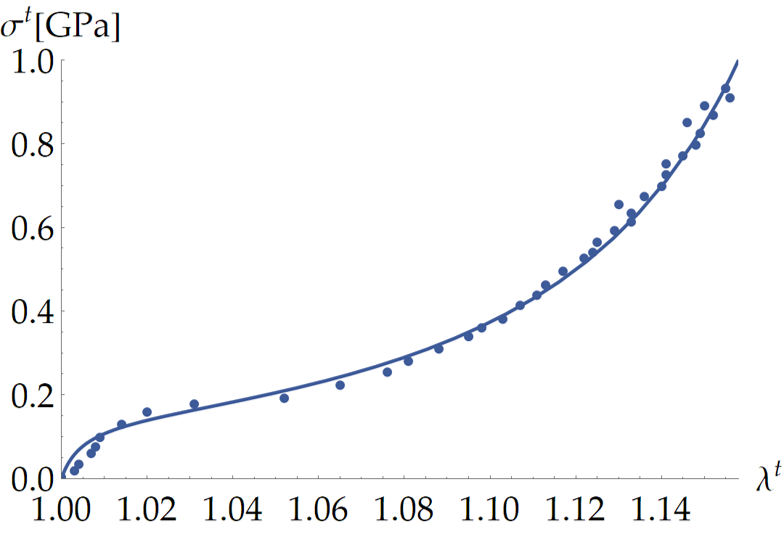} \vspace{-0.15 cm}
	\caption{\label{fig:gosline} Theoretical (continuous lines) {\it vs} experimental (dots) stress-stretch curves for {\it Araneus diadematus} spider fibers (reproduced from \cite{gosline_1999}). 			Here  $
		k^h=6.75 $~GPa, $
		k^s=13.6 $~MPa, $
		\mu=1$~MPa, $
		c_1=1.005,\ 
		c_2=1.979,$
		$\alpha=0,\ 
		m_f/m_o=0.13,\ 
		s=4.5,\ 
		\lambda^s_c=1.25, \ \text{\small RH}_c=84\%.$}
\end{figure}

As a last comparison with experimental results, in Fig. \ref{fig:gosline} we show the possibility of predicting with remarkable accuracy the complex mechanical response of a dragline silk in dry condition  (reproduced from \cite{gosline_1999}). Notice that here, the thread is produced by a third species of spider, the {\it Araneus diadematus}.

\end{document}